\documentstyle[12pt,epsfig]{article} 
\begin{document}
\title{Quantum nondemolition measurements of a particle in electric 
and gravitational fields.}
\author{ A. Camacho  $^{\circ}$
\thanks{email: acamacho@aip.de} 
\thanks{Current address: Instituto Nacional de Investigaciones Nucleares, M\'exico}~~and A. Camacho--Galv\'an $^{\diamond}$\thanks{email: abel@servidor.unam.mx}\\
$^{\diamond}$DEP--Facultad de Ingenier{\'\i}a,\\
Universidad Nacional Aut\'onoma de M\'exico, \\ 
$^{\circ}$Astrophysikalisches Institut Potsdam, \\
An der Sternwarte 16, D--14482 Potsdam, Germany.}
\date{}
\maketitle

\begin{abstract}

In this work we obtain a nondemolition variable for the case in which 
a charged particle moves in the electric and gravitational fields of a sphe\-ri\-cal body. 
Afterwards we consider the continuous monitoring of this nondemolition parameter, 
and calculate, along the ideas of the so called restricted path integral formalism, 
the corresponding propagator. Using these results the pro\-ba\-bi\-lities 
associated with the possible measurement outputs are evaluated. The limit of our 
results, as the resolution of the measuring device goes to zero, is analyzed, and 
the dependence of the corresponding propagator upon the strength of the electric and gravitational fields 
is commented. The role that mass plays in the corresponding results, and its possible connection 
with the equivalence principle at quantum level, are studied.
\end{abstract}
\bigskip
\bigskip
\bigskip

\newpage
\section{Introduction}
\bigskip

One of the fundamental conceptual difficulties in modern physics comprises the so called quantum 
measurement problem, which besets quantum theory since its very first days. The quest 
for a solution embraces already many different ideas and mo\-dels, 
and, of course, each one of them claims to be the correct one [1]. This last remark implies that all these 
proposals must be confronted against the experiment.

Though some of the current solutions of this conundrum are, mathematically, e\-qui\-valent [2], 
in this work we will employ (because it allows us to evaluate in a simpler manner propagators) 
the so called restricted path integral formalism (RPIF) [3]. 
This idea explains a continuous quantum measurement with the introduction of a restriction on 
the integration domain of the corresponding path integral. We may reformulate this 
condition in terms of a weight functional (which contains all the information 
about the measuring process), the one has to be considered in the path integral. The introduction 
of this weight functional allows us to analyze the interaction between measuring 
device and measured system without having to consider a particular measuring scheme. 

Another advantage of RPIF lies in the fact that it may be employed to analyze quantum demolition measurements (QDM), 
or to study the so called quantum nondemolition measurement (QNDM) regime [3]. This last kind of measuring process allows the extraction of the necessary information 
from a quantum system with arbitrarily small error, i.e., there is no limit on 
the measurability of the monitored variable [4].

Concerning the possibility of confronting the results of QNDM against experimental ouputs 
it is noteworthy to comment that there are already some results which report back--action 
evasion measurements in connection with optical fields [5]. The possibility of measuring displacements 
of macroscopic mechanical systems within the quantum regime seems very promising [6], where the relevance of 
these kind of experiments is enhanced by the fact that they could be the begining of other projects, for instance, the production of 
nonclassical states from single macroscopic degrees of freedom. 

Another direction in which the ideas behind QNDM could be tested comprises the case of a particle moving in the Earth's 
gravitational field. In this context, there are already some results, not only in connection with QNDM 
[7], but also in relation with QDM [8]. Nevertheless, more work is needed, because the current models assume, from the very begining, 
that the involved particle shows only gravitational interaction. The possible influence, for instance, 
of the Earth's electric field is neglected. 

In this work we will consider a more realistic scenario, i.e., the case of a particle moving in a region in which the 
electric and gravitational fields of a spherical body are present. The condition 
that defines a quantum nondemolition measurement will be analzyed and solved. 
The co\-rres\-ponding propagator, when this parameter is being continuously monitored, 
will be calculated, and the probabilities, associated with the possible measurement outputs, will 
be obtained. 
The dependence of the corres\-ponding propagator upon the strength of the electric and gravitational fields 
is also commented. 
\bigskip

\section{Propagators}
\bigskip

Consider a spherical body whose radius, mass, and electric charge are $R$, $M$, 
and $Q$, respectively. We assume that this body has no rotation, and for the sake of 
simplicity it will be also supposed that this body has a constant electric charge density. The 
coordinate system will be chosen such that its origin coincides with the center of this body. 

Let us now introduce in this scheme a particle, whose mass and charge are $m$ and 
$q$, respectively. Under these conditions the Lagrangian of this particle reads
\bigskip

{\setlength\arraycolsep{2pt}\begin{eqnarray}
L = {\vec {P}^2\over2m} + G{Mm\over r} - k{qQ\over r}.
\end{eqnarray}
\bigskip

Write now $r = R + z$, where $z$  is
the distance above the body's surface. Under the condition $R>>z$ we may approximate 
the Lagrangian, up to second order in $z$, as follows
\bigskip

{\setlength\arraycolsep{2pt}\begin{eqnarray}
L = {\vec {P}^2\over2m} + \left(G{Mm\over R} - k{qQ\over R}\right)\left(1 - {z\over R} + {z^2\over R^2}\right).
\end{eqnarray}
\bigskip

If the particle goes from point $N$ to point $W$ (whose coordinates are $(x_N, y_N, z_N)$ 
and $(x_W, y_W, z_W)$, respectively), then, quantum mechanically, it can 
be described by the following propagator

{\setlength\arraycolsep{2pt}\begin{eqnarray}
U(W,\tau'';N, \tau') = \left({m\over 2\pi i\hbar T}\right)
\exp\left\{{im\over 2\hbar T}\left[\left(x_W - x_N\right)^2 + \left(y_W - y_N\right)^2\right]\right\}\nonumber\\
\times\exp\left\{{-imTR^2\omega^2\over 2\hbar}\right\}\int d[z(t)]\exp\left\{{i\over\hbar}\int_{\tau'}^{\tau''}
\left[{m\over 2}\dot{z}^2 + Fz - {m\over 2}\omega^2z^2\right]dt\right\},
\end{eqnarray}}
\bigskip

\noindent here $\sqrt{(x_W - x_N)^2 + (y_W - y_N)^2}$ denotes the projection on the body's 
surface of the distance between $W$ and $N$, and $T = \tau'' - \tau'$. We also have 
introduced two definitions
\bigskip

{\setlength\arraycolsep{2pt}\begin{eqnarray}
F = k{qQ\over R^2} - G{Mm\over R^2},
\end{eqnarray}
\bigskip

{\setlength\arraycolsep{2pt}\begin{eqnarray}
\omega^2 = 2\left(k{qQ\over mR^3} - G{M\over R^3}\right).
\end{eqnarray}
\bigskip

Therefore the problem reduces to the calculation of the path integral of a driven harmonic oscilator, 
which is an already known case [9].

Nevertheless, at this point it is important to distinguish three different situations: 
(i) $kqQ - GMm >0$, (ii) $kqQ - GMm = 0$, and finally (iii) $kqQ - GMm <0$. The first one defines 
the usual one--dimensional oscillator ($\omega^2 > 0$), the second condition corresponds to the case of a free 
particle, and the last one is associated with a 
harmonic oscillator which has a complex frequency ($\omega^2 < 0$).
Clearly, if $q$ and $Q$ have opposite signs, then the corresponding frequency remains 
always in the third case.

For the sake of completeness we present the two involved propagators (the case $kqQ - GMm = 0$ is trivial one).
\bigskip
\bigskip

For $\omega^2 > 0$ (this situation appears if $qQ > {GMm\over k}$, and it is possible only for charges 
with the same sign) we have

{\setlength\arraycolsep{2pt}\begin{eqnarray}
U(W,\tau'';N, \tau') = \tilde{U}(W,\tau'';N, \tau')\left({m\over 2\pi i\hbar T}\right)^{{1\over2}}
\sqrt{{\omega T\over \sin\left(\omega T\right)}}\nonumber\\
\times\exp\left\{{im\omega\over 2\hbar \sin\left(\omega T\right)}\left[\left(z^2_W + z^2_N\right)\cos\left(\omega T\right)- 2z_Wz_N\right]\right\}\nonumber\\
\times\exp\left\{-{imR^2\omega^2\over 4\hbar}
\left[{3T\over 2} + {1 - \cos\left(\omega T\right)\over\omega\sin\left(\omega T\right)}\left({R - 2(z_W + z_N)\over R}\right)\right]\right\},
\end{eqnarray}}
\bigskip

where 

{\setlength\arraycolsep{2pt}\begin{eqnarray}
\tilde{U}(W,\tau'';N, \tau') = \left({m\over 2\pi i\hbar T}\right)\exp\left\{{im\over 2\hbar T}\left[\left(x_W - x_N\right)^2 + \left(y_W - y_N\right)^2\right]\right\}.
\end{eqnarray}}
\bigskip

In the case $\omega^2 < 0$ (which is always the situation if $qQ \leq 0$) 

{\setlength\arraycolsep{2pt}\begin{eqnarray}
U(W,\tau'';N, \tau') = \tilde{U}(W,\tau'';N, \tau')\left({m\over 2\pi i\hbar T}\right)^{{1\over2}}
\sqrt{{|\omega| T\over \sinh\left(|\omega| T\right)}}\nonumber\\
\times\exp\left\{{im|\omega|\over 2\hbar \sinh\left(|\omega| T\right)}\left[\left(z^2_W + z^2_N\right)\cosh\left(|\omega| T\right)- 2z_Wz_N\right]\right\}\nonumber\\
\times\exp\left\{{imR^2|\omega|^2\over 4\hbar}
\left[{3T\over 2} + {\cosh\left(|\omega| T\right) - 1\over\omega\sinh\left(|\omega| T\right)}\left({R + 2(z_W + z_N)\over R}\right)\right]\right\},
\end{eqnarray}}
\bigskip

\noindent here $|\omega| = \sqrt{2|\left(k{qQ\over mR^3} - G{M\over R^3}\right)|} \in \Re$.
\bigskip

We may notice that the case $qQ < 0$ renders a frequency 
which has a larger magnitude than the corresponding magnitude when $k = 0$. Indeed in this situation $|\omega| = \sqrt{2\left(G{M\over R^3} + k{|qQ|\over mR^3}\right)} > \sqrt{2G{M\over R^3}}$. 
On the other hand, if the charges have the same sign, 
then the frequency becomes smaller, i.e., $|\omega| = \sqrt{2\left(G{M\over R^3} - k{|qQ|\over mR^3}\right)} < \sqrt{2G{M\over R^3}}$.
\bigskip

\section{Continuous quantum measurements}
\bigskip

\subsection{Path integrals and quantum measurements}
\bigskip

Let us now introduce a measuring process, namely we will monitor, continuously, 
some parameter associated to the description of the involved particle. 
According to RPIF [3] this kind of measuring process can be described by the introduction of a weight functional in the corresponding 
path integral, this weight functional contains all the information concerning 
the interaction between measuring device and measured system, in other words, all the information about the measuring process. An advantage 
of this formalism lies in the fact that we do not need consider the introduction 
of a particular measuring scheme.

Nevertheless, we face, at this point, a problem, namely in order to obtain theo\-re\-tical predictions, 
we must choose a par\-ti\-cu\-lar expression for the involved weight functional.

Our choice will be a gaussian weight functional, and a justification for it stems from 
the fact that the results coming from a Heaveside weight functional [10] and those 
coming from a gaussian one [11] coincide up to the order of magnitude. Hence, 
if in this first approach we are interested in the order of magnitude of the involved 
effects, then we may assume this approximation. It can be supposed that we have an 
experimental device whose weight functional has precisely this form. 
We may wonder if this is not an unphysical assumption, and in favor of this argument 
it can be commented that recently it has been proved 
that there are measuring processes in which the corresponding weight functional 
behaves precisely in this manner [12].

This means that if we measure, continuously, the observable $A$ of our particle, 
then we  may introduce in the involved path integral the following weight functional
\begin{equation}
\omega_{[a(t)]}[A(t)] = \exp\left\{-{1\over T\Delta a^2}\int _{\tau '}^{\tau ''}[A(t) - a(t)]^2dt\right\},
\end{equation}
\bigskip

\noindent here $\Delta a$ represents the error in our measurement (the resolution 
of the measuring device), while function $a(t)$ is the measurement output.
The squared modulus of the complete propagator (including (9)) allows us to evaluate 
the probability of obtaining as measurement output function $a(t)$.
\bigskip

\subsection{Demolition measurements}
\bigskip

The monitoring of position is a demolition measurement, this phrase means that in 
these kind of measuring processes the corresponding observable has always a\-sso\-cia\-ted 
an absolute limit on its measurability. This restriction is a consequence of the unavoidable 
back reaction of the measuring device upon the measured system and stems 
from Heisenberg uncertainty relations [4]. 

The case of position monitoring of a particle moving in the gravitational field of an spherical body 
has already been analyzed [8]. In the present situation, when $\omega^2 < 0$, the only modification 
that has to be introduced (with respect to the corresponding expressions in [8]), comprises the change of two  
parameters, namely the driving force and the frequency become now
$F(t) = k{qQ\over R^2}-G{Mm\over R^2} - {4i\hbar z(t)\over T\Delta z^2}$ 
and $\omega^2 = 2\left(k{qQ\over R^3} - G{M\over R^3}\right) - {4i\hbar\over mT\Delta z^2}$, respectively. 
It is readily seen that the electric field modifies solely the real part of $F(t)$ and 
$\omega^2$. This last remark is a consequence of the fact that the imaginary parts 
of driving force and frequency are determined only by the interaction with the measuring 
device.

The case $\omega^2 > 0$ renders a harmonic oscillator whose coordinate is being monitored, 
and its propagator is also already known [3].
\bigskip

\bigskip

\subsection{Nondemolition measurements}
\bigskip

The idea behind a quantum nondemolition measurement is to monitor a variable such that 
the unavoidable disturbance of the complementary parameter (stemming 
from Heisenberg uncertainty relations) does not disturb the evolution 
of the chosen variable [4]. This implies that in these type of measuring processes 
there is no absolute limit on the measurability of the monitored variable, i.e., 
we may extract the necessary information about the involved quantum system with an 
arbitrarily small error.
\bigskip

The Hamiltonian related to (2)

\begin{equation}
H = {\left(p_x^2 + p_y^2 + p_z^2\right)\over 2m} - F(t)z + {m\omega^2\over 2}z^2.
\end{equation}
\bigskip

Let us now suppose that we measure, continuously, the observable $\Gamma$, and 
that as measurement output this experiment renders function $\gamma(t)$. 

Hence the propagator of the involved particle is now

{\setlength\arraycolsep{2pt}\begin{eqnarray}
U_{[\gamma(t)]}(W, \tau''; N, \tau') = \left({m\over 2\pi i\hbar T}\right)
\exp\left\{{im\over 2\hbar T}\left[\left(x_W - x_P\right)^2 + \left(y_W - y_P\right)^2\right]\right\}\nonumber\\
\times\exp\left\{i{mTR^2|\omega|^2\over 2\hbar}\right\}\int d[z]d[p]
\exp\left\{{i\over\hbar}\int_{\tau '}^{\tau ''}\left[{p^2_z\over 2m} + F(t)z - {m\omega^2\over 2}z^2\right]dt\right\}\times\nonumber\\
\exp\left\{-{1\over T\Delta \gamma^2}\int _{\tau '}^{\tau ''}\left[\Gamma(t) - \gamma(t)\right]^2dt\right\}.
\end{eqnarray}}
\bigskip

We have the case of a harmonic oscillator, with the additional complication that 
if  $\omega^2 < 0$, then the frequency becomes complex. Hence we may (as in done 
in the case $\omega^2 > 0$ [13]) consider that

\begin{equation}
\Gamma = \theta z + \psi p_z,
\end{equation}
\bigskip

\noindent where $\theta, \psi: \Re\rightarrow\Re$.
\bigskip

The differential equation that determines when $\Gamma$ defines a quantum 
nondemolition variable reads [3]

\begin{equation}
{d\over dt}\left({\theta\over\psi}\right)= {1\over m}\left({\theta\over\psi}\right)^2 + m\omega^2.
\end{equation}
\bigskip

If $\omega^2 > 0$ we have

\begin{equation}
{\theta\over\psi} = m\omega\tan{(\omega t)}.
\end{equation}
\bigskip

The differential equation, if $\omega^2 < 0$, can also be solved analytically

\begin{equation}
{\theta\over\psi} = -m|\omega|\tanh{(|\omega| t)}.
\end{equation}
\bigskip

Hence, a nondemolition variable is

\begin{equation}
\Gamma = p_z - m|\omega|z\tanh{(|\omega| t)},
\end{equation}
\bigskip

\noindent here we have assumed, for the sake of simplicity, that $\psi = 1$.

Employing this last expression in (11), we obtain the new propagator, which is of 
gaussian type, and therefore can be calculated. Hence, the probability density
of obtaining $\gamma(t)$ as measurement output is given, after a lengthy calculation, by $P_{[\gamma(t)]} = |U_{[\gamma(t)]}|^2$. In our case 
this expression is
\bigskip

{\setlength\arraycolsep{2pt}\begin{eqnarray}
P_{[\gamma(t)]} = {m\over 2\pi\hbar}\exp\left\{-{2\over T\Delta\gamma^2}\left[1 + 
{4m^2\hbar^2\over 4m^2\hbar^2 + T^2\Delta\gamma^4}\right]\int _{\tau '}^{\tau ''}\gamma^2dt 
-{4F\over T\Delta \gamma^2}\int _{\tau '}^{\tau ''}\tilde{\gamma}dt\right\}\nonumber\\
\times\sqrt{\Omega^2 + \Pi^2\over\nu}\exp\left\{{8\hbar^2\over T\Delta\gamma^2\left[4m^2\hbar^2 + T^2\Delta\gamma^4\right]}
\int _{\tau '}^{\tau ''}\tilde{\gamma}f\left[2m\gamma + \tilde{\gamma}f\right]dt\right\}\nonumber\\
\times\exp\left\{{m\over\nu\hbar}\left(z^2_W + z^2_N\right)A_1 + 2{z_W + z_N\over\nu\hbar\left[\Omega^2 + \Pi^2\right]}A_2\right\}\nonumber\\
\times\exp\left\{2m{z_Wz_N\over\nu\hbar}A_3 + 2{T\over m\hbar}A_4 + 
2{\cosh(\Pi T) - \cos(\Omega T)\over m\nu\hbar}A_5\right\}.
\end{eqnarray}}
\bigskip

The definitions introduced in this last expression appear, for the sake of clarity, in the appendix.

\subsection{Conclusions}
\bigskip

In this work a nondemolition variable for the case of a charged particle moving in the 
electric and gravitational fields of a spherical body has been obtained. This 
particle has been subject to a continuous monitoring of this nondemolition parameter, 
and using RPIF, not only its corresponding path integral has been calculated, but 
also the probabilities associated with the different measurement outputs have been evaluated. 

It has been found that if the resolution of the measuring device 
becomes much smaller than the quantum threshold of the system, $m\hbar/T$, 
then all the possible measurement outputs have the same probability. 
This equiprobability of all the possible measurement outputs is clearly a 
quantum feature.

Another interesting feature of expressions (18--20) comprises the role that mass plays at 
quantum realm. At this point it is noteworthy to mention that in the case in which 
position is being monitored [8, 9], mass always appears in the combination $m/\hbar$. 
For a nondemolition variable this is not true, for instance, in the first exponential in (18) we have the combination 
$m^2\hbar^2$.

Once again mass appears explicitly in the expression for probabilities, a fact that 
seems to support the conclusion that at quantum level gravity is not purely geometric [14], 
i.e., according to (18--20) mass effects depend in a complicated manner upon $m$. 
Concerning this last remark it is noteworthy to mention that the quantum threshold of the system also depends explicitly upon the mass of the involved particle. 
This is an obvious statement, but it implies that in a gravitational field 
two particles, with different masses, do not have the same quantum threshold. 
Clearly the geometrization of any mass dependent concept could be a hard task, 
and therefore it seems that the geometrization of the concept of quantum 
threshold (the one tells us when quantum noise plays an important 
role in a measuring process) could face conceptual difficulties. 
At this point we could ask, which concepts (if any), stemming from quantum measurement 
theory, could be incorporated as geometric entities in general relativity. From 
our argumentation it seems that, at least in the contex of the present work, 
quantum threshold is not a viable candidate. 

According to general relativity, around any point in any curved manifold, a 
neighborhood can be found, in which the laws of physics are those valid in a 
Minkowskian spacetime [15]. Recently it has been claimed that there is an incompatibility 
between quantum measurement theory and special relativity [16], in the sense 
that quantum measurement requires a preferred Lorentz frame. Hence, joining the 
last two co\-mments we could be naive enough and
extrapolate this last claim and wonder if quantum measurement theory could 
also be incompatible with general relativity. The argumentation around 
the viability of the geometrization of our quantum threshold concept seems to 
point in the direction of this incompatibility.
Nevertheless, this topic and the possible violation of the 
equivalence principle at quantum level are cu\-rrent\-ly controversial issues 
[17, 18, 19, 20], and of course, more work is needed in this direction.
\bigskip 

\subsection{Appendix}
\bigskip

Here we give the definitions that have been introduced in expression (17).

{\setlength\arraycolsep{2pt}\begin{eqnarray}
f = -m\omega\tanh\left(|\omega|t\right),
\end{eqnarray}}

{\setlength\arraycolsep{2pt}\begin{eqnarray}
\hat{f}^2 = {1\over T}\int _{\tau '}^{\tau ''}f^2(t)dt,
\end{eqnarray}}

{\setlength\arraycolsep{2pt}\begin{eqnarray}
\tilde{\gamma}(t) = \int _{\tau '}^{t}\gamma(\tau)d\tau,
\end{eqnarray}}

{\setlength\arraycolsep{2pt}\begin{eqnarray}
\Omega = \sqrt{{\sqrt{m^2\omega^4T^2\Delta\gamma^4 + 4\hbar^2(m^2\omega^2 + 
\hat{f}^2)^2}\over mT\Delta\gamma^2}}\cos\left\{{1\over2}\arctan\left({2\hbar[\hat{f}^2 + m^2\omega^2]
\over m\omega^2T\Delta\gamma^2}\right)\right\},
\end{eqnarray}}
\bigskip

{\setlength\arraycolsep{2pt}\begin{eqnarray}
\Gamma = \sqrt{{\sqrt{m^2\omega^4T^2\Delta\gamma^4 + 4\hbar^2(m^2\omega^2 + 
\hat{f}^2)^2}\over mT\Delta\gamma^2}}\sin\left\{{1\over2}\arctan\left({2\hbar[\hat{f}^2 + m^2\omega^2]
\over m\omega^2T\Delta\gamma^2}\right)\right\},
\end{eqnarray}}

{\setlength\arraycolsep{2pt}\begin{eqnarray}
\nu = \sin^2(\Omega T)\cosh^2(\Pi T) + \sinh^2(\Pi T)\cos^2(\Omega T),
\end{eqnarray}}

{\setlength\arraycolsep{2pt}\begin{eqnarray}
A_1 = \Omega\sinh(\Pi T)\cosh(\Pi T) - \Gamma\sin(\Omega T)\cos(\Omega T ),
\end{eqnarray}}

{\setlength\arraycolsep{2pt}\begin{eqnarray}
\varphi = \arctan\left(-{2m\hbar\over T\Delta\gamma^2}\right),
\end{eqnarray}}

{\setlength\arraycolsep{2pt}\begin{eqnarray}
\hat{F} = \left[1 + {4m^2\hbar^2\over T^2\Delta\gamma^4}\right]^{1/4}
\left\{F + {4\hbar^2\hat{f}^2<\tilde{\gamma}>\over 4m^2\hbar^2 + T^2\Delta\gamma^4} 
-4{m\hbar^2\hat{f}<\gamma>\over 4m^2\hbar^2 + T^2\Delta\gamma^4}\right\}\cos\left({\varphi\over 2}\right)\nonumber\\
-\left[1 + {4m^2\hbar^2\over T^2\Delta\gamma^4}\right]^{1/4}
\left\{{2\hbar\hat{f}\left[T^2\Delta\gamma^4<\gamma> - 4\hbar^2m\hat{f}<\tilde{\gamma}>\right]\over T\Delta\gamma^2\left[4m^2\hbar^2 + T^2\Delta\gamma^4\right]}
\right\}\sin\left({\varphi\over 2}\right)\nonumber\\
+\left[1 + {4m^2\hbar^2\over T^2\Delta\gamma^4}\right]^{1/4}\left\{{2m\omega^2\hbar<\tilde{\gamma}>\over T\Delta\gamma^2}\right\}\sin\left({\varphi\over 2}\right),
\end{eqnarray}}

{\setlength\arraycolsep{2pt}\begin{eqnarray}
\tilde{F} = \left[1 + {4m^2\hbar^2\over T^2\Delta\gamma^4}\right]^{1/4}
\left\{F + {4\hbar^2\hat{f}^2<\tilde{\gamma}>\over 4m^2\hbar^2 + T^2\Delta\gamma^4}
+4{m\hbar^2\hat{f}<\gamma>\over 4m^2\hbar^2 + T^2\Delta\gamma^4}\right\}\sin\left({\varphi\over 2}\right)\nonumber\\
+\left[1 + {4m^2\hbar^2\over T^2\Delta\gamma^4}\right]^{1/4}
\left\{{2\hbar\hat{f}\left[T^2\Delta\gamma^4<\gamma> - 4\hbar^2m\hat{f}<\tilde{\gamma}>\right]\over T\Delta\gamma^2\left[4m^2\hbar^2 + T^2\Delta\gamma^4\right]}\right\}\cos\left({\varphi\over 2}\right)\nonumber\\
-\left[1 + {4m^2\hbar^2\over T^2\Delta\gamma^4}\right]^{1/4}\left\{{2m\omega^2\hbar<\tilde{\gamma}>\over T\Delta\gamma^2}\right\}\cos\left({\varphi\over 2}\right),
\end{eqnarray}}

{\setlength\arraycolsep{2pt}\begin{eqnarray}
A_2 = \left[\cosh(\Pi T) - \cos(\Omega T)\right]
\left[\sin(\Omega T)(\hat{F}\Pi - \tilde{F}\Omega) - \sinh(\Pi T)(\hat{F}\Omega + \tilde{F}\Pi)\right],
\end{eqnarray}

{\setlength\arraycolsep{2pt}\begin{eqnarray}
A_3 = \Pi\sin(\Omega T)\cosh(\Pi T) - \Omega\sinh(\Pi T)\cos(\Omega T),
\end{eqnarray}}

{\setlength\arraycolsep{2pt}\begin{eqnarray}
A_4 = {\tilde{F}\hat{F}\left(\Pi^2 - \Omega^2\right) + \Omega\Pi\left(\hat{F}^2 - \tilde{F}^2\right)
\over\left(\Omega^2 +\Pi^2\right)^2},
\end{eqnarray}}

{\setlength\arraycolsep{2pt}\begin{eqnarray}
A_5 = \Omega\left(\Omega^2 - 3\Pi^2\right){2\tilde{F}\hat{F}\sin(\Omega T) + 
\left(\hat{F}^2 - \tilde{F}^2\right)\sinh(\Pi T)\over 
\Omega^2\left(\Omega^2 - 3\Pi^2\right)^2 + \Pi^2\left(\Pi^2 - 3\Omega^2\right)^2}\nonumber\\
\Pi\left(3\Omega^2-\Pi^2\right){2\tilde{F}\hat{F}\sinh(\Pi T) + 
\left(\tilde{F}^2 - \hat{F}^2\right)\sin(\Omega T)\over 
\Omega^2\left(\Omega^2 - 3\Pi^2\right)^2 + \Pi^2\left(\Pi^2 - 3\Omega^2\right)^2}.
\end{eqnarray}}
\bigskip

Here the notation $<>$ means always the time average of the corresponding variable in the interval 
$[\tau', \tau'']$.
\bigskip

\Large{\bf Acknowledgments}\normalsize
\bigskip

A. C. would like to thank A. A. Cuevas--Sosa his help, and D.-E. Liebscher for the fruitful discussions on the subject. 
The hospitality of the Astrophysika\-li\-sches Institut Potsdam is also kindly acknowledged. 
This work was supported partially by CONACYT (M\'exico) Posdoctoral Grant No. 983023.

\end{document}